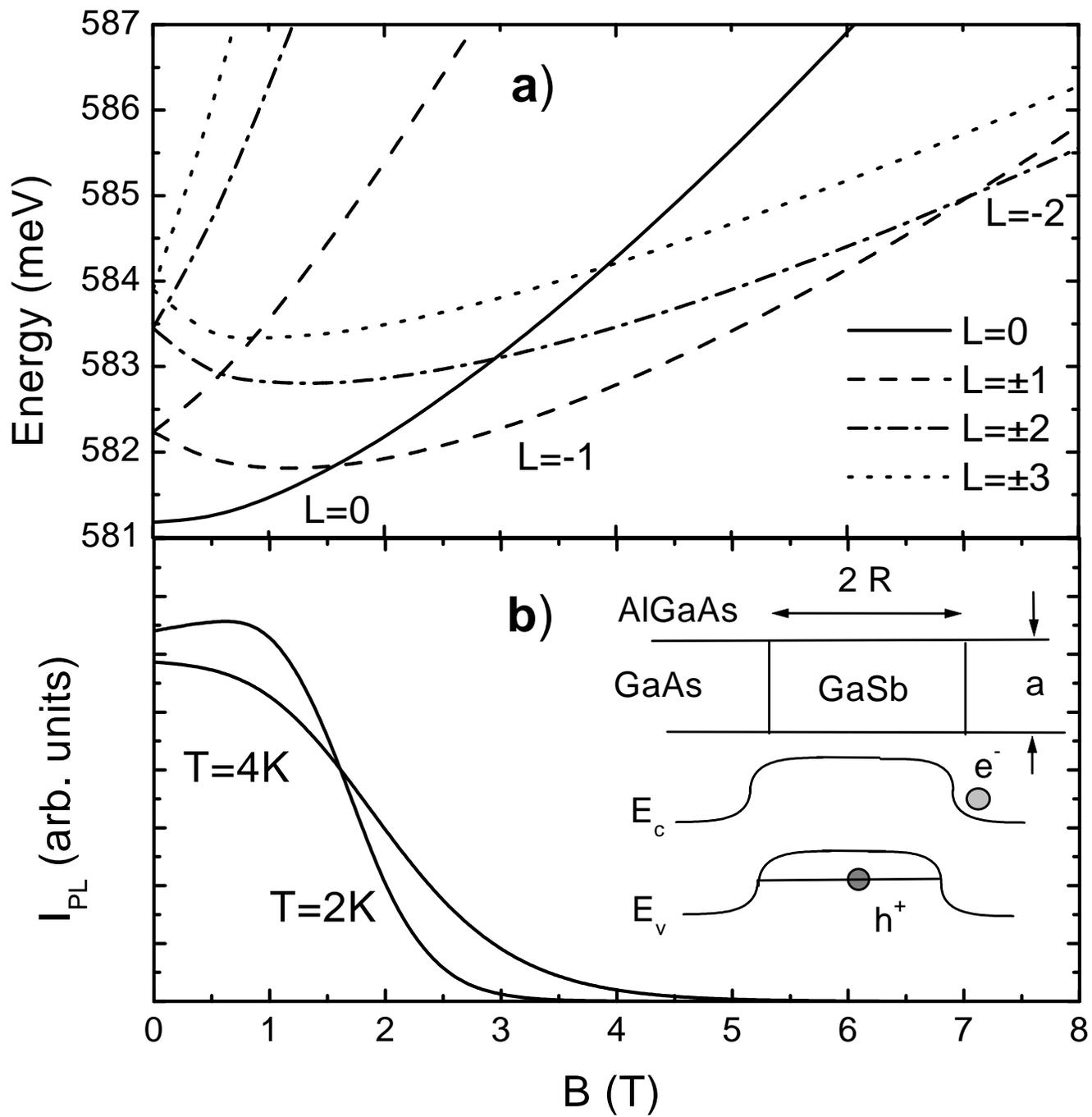

Fig. 1

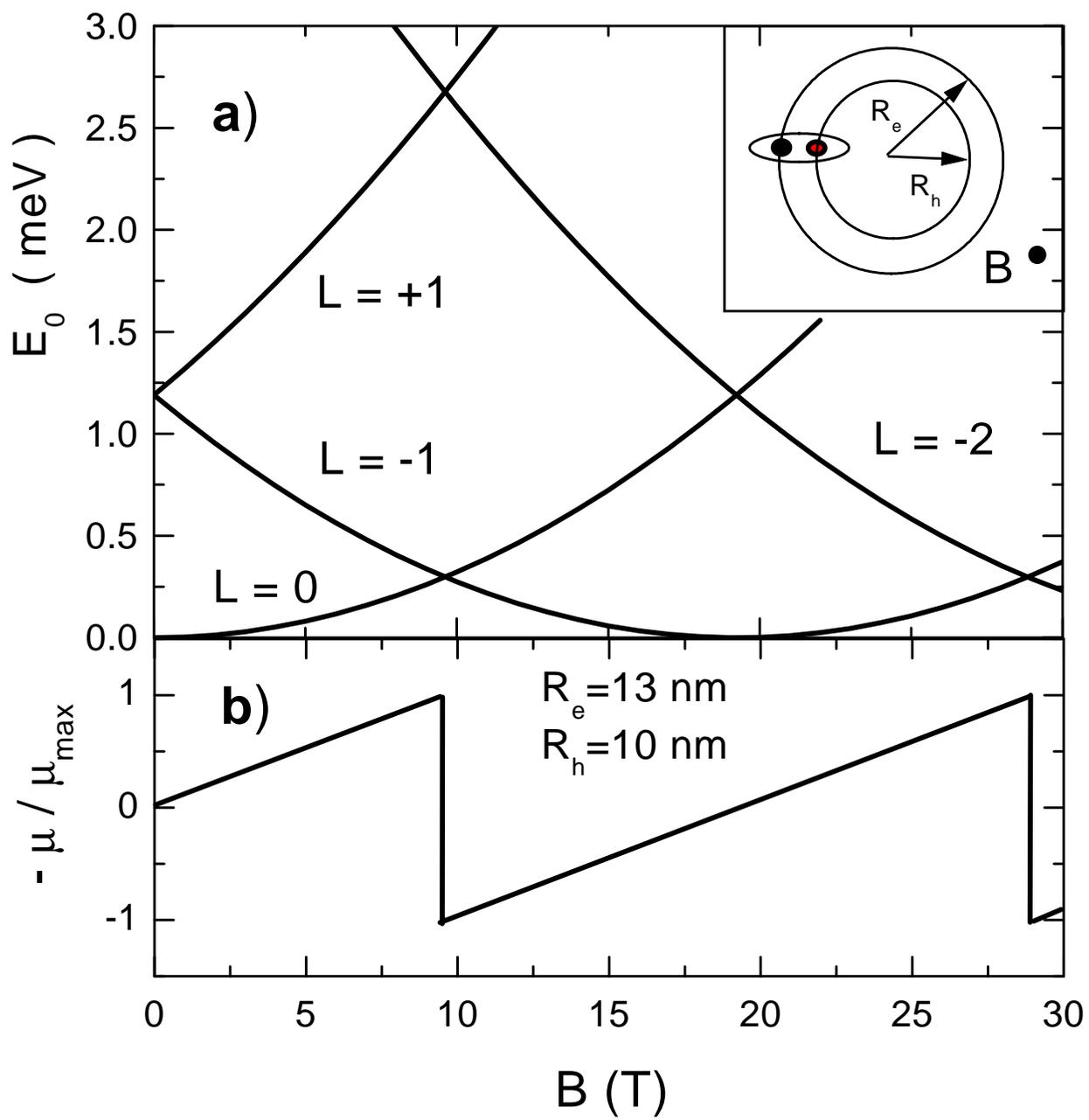

Fig. 2

# Excitons in quantum-ring structures in a magnetic field:

# Optical properties and persistent currents

A.O. Govorov[a,b], A.V. Kalameitsev[b], R. Warburton[c], K. Karrai[d], and S.E. Ulloa[a]

[a] Department of Physics and Astronomy, and CMSS Program, Ohio University, Athens, OH 45701, USA; govorov@helios.phy.ohiou.edu

[b] Institute of Semiconductor Physics, Russian Academy of Sciences, Siberian Branch, 630090 Novosibirsk, Russia

[c] Department of Physics, Heriot-Watt University, Edinburgh - UK

[d] Center for NanoScience, LMU, Geschwister-Scholl-Platz 1, 80539 München, Germany

We study theoretically the magnetic field effect on a neutral, but polarizable exciton confined in quantum-ring structures. For excitons with a nonzero dipole moment, a novel magnetic interference effect occurs: The ground state of an exciton confined in a finite-width quantum ring possesses a nonzero angular momentum with increasing normal magnetic field. This effect is accompanied by a suppression of the photoluminescence in well-defined magnetic-field intervals. The magnetic interference effect is calculated for type-II quantum dots and quantum rings.

To whom correspondence should be addressed: Alexander Govorov, Department of Physics and Astronomy, and CMSS Program, Ohio University, Athens, OH 45701, USA; govorov@helios.phy.ohiou.edu ; fax: ++1-740-593-0433

The growth of ring-shaped semiconductor quantum dots with nm-scale radii has triggered much interest in the theoretical and experimental investigation of their electronic and optical properties [1-5]. Here we consider the effect of static high magnetic fields on the optical spectra of excitons confined in semiconductor quantum rings. The joint action of the Coulomb forces and the shape of the nano-structure potential on electrons and holes make this problem interesting and challenging. At high enough magnetic fields, we found that the ground state of a neutral magneto-exciton confined in a finite-width quantum ring possesses a nonzero angular momentum. This effect can be observed as a suppression of the photoluminescence (PL) in well-defined intervals of the magnetic field. Recent predictions show that the binding energy of a neutral exciton confined in a one-dimetional (1D) ring oscillates with increasing magnetic field [3]. This Aharonov-Bohm (AB) effect for a neutral exciton occurs due to electron-to-hole tunneling around the ring [3]. However, the amplitude of the predicted AB oscillations in the exciton binding energy is very sensitive to a finite width of a quantum ring [4] and includes an exponentially-small tunneling amplitude. Here, we describe a different magnetic interference effect, which relates to the polarized exciton in a quantum ring with a finite width and does _not_ include an exponentially-small tunneling factor.

Usually, the confining potentials in a ring-shaped semiconductor quantum dots are different for electrons and holes [2]. As a result of this, the potential can polarize a quantum-ring exciton in the radial direction. When a dipole moves along a closed trajectory in a ring, its wave function acquires a nonzero magnetic phase and the quantum interference occurs. In order to demonstrate the magnetic-field effect, we present two models. The first model relates to a type-II quantum dot embedded in a 2D quantum well [5,6]. This type-II potential would correspond to that of GaSb/GaAs quantum dots [7]. The second model describes the exciton spectrum in the InAs self-organized quantum rings [1,2].

***Type-II quantum dots.*** In a type-II electron-hole system, the potential of a quantum dot localizes a hole near the center whereas an electron moves in a quantum-ring potential due to the joint action of the Coulomb force and the quantum-dot potential (insert of Fig. 1). Here we model a GaSb-GaAs quantum dot as a GaSb cylinder embedded into a 2D quantum well. In contrast to our previous calculations on magneto-excitons in type-II QD's, here we include the effect of smooth interfaces and calculate the PL intensity at different temperatures. In such a model, the motion in the normal z-direction is strongly quantized and can be separated from the in-plane motion. Another convenient simplification occurs in the case of a strong confinement of a hole: the electron-hole Coulomb potential weakly perturbs the hole motion. At the same time, the Coulomb potential is very important in the description of the electron motion, since it results in electron localization around a ring-like potential near the quantum dot (inset of Fig. 1). Also, we assume that the electron (hole) potentials, $U_{e(h)}(\rho)$, have cylindrical symmetry. Here $\rho$ is the distance to the quantum-dot center. By using the above conditions we can write the in-plane wave function of exciton in the form $\Psi = e^{iL_e\phi_e} e^{iL_h\phi_h} f_e(\rho_e) f_h(\rho_h)$, where $(\phi_{e(h)}, \rho_{e(h)})$ are the in-plane electron (hole) coordinates, $f_{e(h)}$ are the radial wave functions, and $L_{e(h)}$ are the angular momenta. This wave function includes correlations between the radial motions of the particles [8], as we now explain. The in-plane radial electron motion is given by the Hamiltonian:

$$\hat{H}_e = -\frac{\hbar^2}{2m_e}\left[\frac{\partial^2}{\partial \rho_e^2} + \frac{1}{\rho_e}\frac{\partial}{\partial \rho_e} - \frac{L_e^2}{\rho_e^2}\right] + \frac{\hbar\omega_{ce}}{2}L_e + \frac{m_e\omega_{ce}^2}{8}\rho_e^2 + U_e(\rho_e) + U_C(\rho_e),$$

(1)

where $m_e$ is the electron effective mass, $\omega_{ce}$ is the electron cyclotron frequency, and $U_C$ is the Coulomb potential due to the localized hole. The hole wave function is mostly determined by the nanostructure potential $U_h(\rho)$ and weakly depends on the magnetic field because of strong localization. In the following, we consider only the ground state of the hole with $L_h = 0$. For the electron states, the wave function has two quantum numbers: $L_e$ and the radial number $n_e = 0, 1, 2, \ldots$. Then, the exciton energies can be written in the form $E_{exc}(L, n_e) = E_e(L_e, n_e) + E_h + E_g$, where $L = L_h + L_e$ is the total angular momentum, $E_e$ is the electron energy given by operator (1), $E_h$ is the hole energy, and $E_g$ is the bandgap energy for the indirect heterojunction GaAs-GaSb. To include the effect of atomic interdiffusion at the GaAs-GaSb interface, we introduce a smoothing function for the in-plane band mismatch with a characteristic length $r_d$. The QD diameter is assumed to be $d = 2R = 10 nm$. For the calculations, we used $r_d = 0.55 nm$, the quantum well width $a = 10 nm$, and band-structure parameters of the GaAs-GaSb system.

Our calculations show that the ground state momentum of an exciton changes from $L = 0$ to $L = -1, -2, -3, \ldots$ as the magnetic field $B$ increases (Fig. 1a). This is due to the ring-like motion of electron in the polarized exciton confined in a type-II QD. According to the selection rules for optical inter-band transitions, only the exciton with $L = 0$ can emit a photon. Also, at low temperatures, the PL comes mostly from the exciton ground state due to fast exciton relaxation. This means that the PL signal from the type-II QD becomes suppressed when the magnetic field induces the transition $L = 0 \to L = -1$. Such a behavior is clearly seen in the calculated PL intensity $I_{PL}$ at low temperature (Fig. 1b). For our calculations we used the relation: $I_{PL} \propto A^2 P_{l=0}(T)$, where $P_{l=0}(T)$ is the probability of exciton being in the state with $L=0$ and $n_e=0$ at temperature $T$, and $A$ is the electron-hole overlap integral. In our model, the integral $A$ first slowly increases with magnetic field when the field is relatively weak. This is due to additional localization of the electron around the type-II QD. At very high magnetic fields, however, when the hole becomes strongly localized in the center of a dot, the overlap integral starts to decrease with magnetic field. The calculated magnetic-field dependence of $I_{PL}$ comes mainly from the probability factor $P_{l=0}$ and has a peculiar $T$-dependence: with decreasing temperature the PL intensity $I_{PL}(B)$ approaches a step-like function. Such $T$-dependence can be used to recognize the magnetic-field-interference mechanism described here.

*Quantum rings (QR's).* In self-organized QR structures, the electron and hole potentials for the in-plane motion can be locally approximated by $U_{e(h)} = m_{e(h)} \Omega_{e(h)}^2 (\rho - R_{e(h)})^2 / 2$ [1], where $\Omega_{e(h)}$ represent the characteristic radial oscillation frequencies and $R_{e(h)}$ are the ring radii for electron and hole. Here we like to note that the potentials $U_{e(h)}$ are <u>not</u> harmonic as they are defined for $\rho \geq 0$. In the vertical, $z$-direction, the motion is assumed to be strongly quantized. First we analyze single-particle wave functions for the case $R_e = R_h$. Since in most cases $m_e \Omega_e \neq m_h \Omega_h$ [2] and the potentials are essentially anharmonic, the single-particle electron and hole wave functions do not coincide. This means that even for the case $R_e = R_h$ the asymmetry in effective masses and potentials results in different magnetic-field dispersions for the electron and hole single-particle energies, $E_{e(h)}^{sp}(B)$. By calculating the wave functions for realistic parameters from Refs. 1-2, we find that the electron and hole wave functions are peaked at different radii and, thus, the exciton has nonzero polarization in the radial direction. The latter results in ground-state transitions with increasing magnetic field. In the following, we will demonstrate this effect using a model of two 1D ring with $R_e \neq R_h$, one for the electron and one for the hole (insert of Fig. 2).

The electron-hole asymmetry in a QR can be strongly enhanced in the presence of an impurity charge in the ring center or due to a metal nano-gate with an applied voltage in the center, resulting in quite different effective radii for the electron and hole, $R_e \neq R_h$. The problem is analytically solved in the case of strong quantization in the radial direction [9]. This allows us to separate variables in the exciton wave function: $\Psi = f_e(\rho_e) f_h(\rho_h) \psi(\phi_e, \phi_h)$, where the functions $f_{e(h)}(\rho)$ describe radial motions. The azimuthal motion along the ring is given by the Hamiltonian: $\hat{H}_{exc} = \hat{T}_e(\phi_e) + \hat{T}_h(\phi_h) + u_C(\phi_e - \phi_h)$, where $\hat{T}_{e(h)}$ are the kinetic energy operators in the presence of the magnetic field, and $u_C$ is the Coulomb potential averaged over the coordinate $\rho$ involving the functions $f_{e(h)}(\rho)$. By introducing new variables in the operator $\hat{H}_{exc}$, we can separate the "center-of-mass" motion in the exciton from the motion related to the internal coordinate $\phi_e - \phi_h$. Details of calculation can be found in Ref. 9.

The limit of strong Coulomb interaction in the exciton implies the condition $R_0 \gg a_0$, where $a_0$ is the effective Bohr radius in the semiconductor and $R_0 = (R_e + R_h)/2$. In this case, the electron and hole motions are strongly correlated and the electron-hole pair rotates along the ring as a whole quasi-particle with energy $E_0 = \hbar^2/(2MR_0^2)[L + \Delta\Phi/\Phi_0]^2$, where $\Phi_0 = hc/e$ and $\Delta\Phi = \pi(R_e^2 - R_h^2)B$ is the magnetic flux penetrating the area between the electron and hole trajectories;

$M = (m_e R_e^2 + m_h R_h^2)/R_0^2$ and $m_{e(h)}$ are effective masses. This rotation energy is responsible for the magnetic-field dependence of the total exciton energy. With increasing magnetic field, the ground state is successively changed from a state with $L = 0$ to states with $L = -1, -2, \ldots$, like in type-II QD's (Fig. 2). The PL intensity also behaves similarly to the case of type-II QD's [9].

Another physical quantity, which oscillates due to the ground-state transitions, is the magnetization associated with the persistent current in an exciton (Fig. 2b). For the magnetic moment of the exciton we obtain: $\mu = -\partial E_0 / \partial B = -2\mu_{max}(L + \Delta\Phi/\Phi_0)$, where $\mu_{max} = \mu_B (m_e/M)(R_e^2 - R_h^2)/2R_0^2$ and $\mu_B$ is the Bohr magneton.

The limit of weak Coulomb interaction corresponds to the case of relatively small rings with $R_0 \ll a_0$. The single-particle rotation energy of exciton in this case has the form: $E_0^{sp} = \varepsilon_e(L_e + \Phi_e/\Phi_0)^2 + \varepsilon_h(L_h - \Phi_h/\Phi_0)^2$, where $\varepsilon_{e(h)} = \hbar^2/2m_{e(h)}R_{e(h)}^2$ and $\Phi_{e(h)} = \pi R_{e(h)}^2 B$. For this regime, the character of transitions is a bit more complicated: The ground state $(L_e, L_h) = (0, 0)$ for the ring with $R_e > R_h$ is changed to the states $(-1, 0), (-1, 1), (-2, 1)$, etc. with increasing magnetic field [9]. Correspondingly, the total momentum is changed from $L = 0$ to $-1, 0, -1, \ldots$. In the magnetic-field windows when the momentum of the ground state $L = L_e + L_h$ becomes nonzero, the low-temperature PL intensity becomes strongly suppressed [9]. By comparing the limits of strong and weak Coulomb interactions, we see that the character of ground-state transitions depends on the strength of electron-hole correlations.

To conclude, we have studied the magnetic interference effects for a neutral exciton localized in quantum-ring systems which originate from the nonzero magnetic flux $\Delta\Phi$ through the area between the electron and hole trajectories (insert of Fig. 2) . The signature of the predicted ground-state transitions and persistent currents in an exciton is darkness of the PL in well-defined intervals of the magnetic field. Single-dot spectroscopy can be a suitable method to observe this effect [2,10].

*Acknowledgements.* We gratefully acknowledge financial support by Ohio University through the Rufus Putnam Visiting Professorship, by the Volkswagen-Foundation, the RFBR (Russia), and the US Department of Energy grant no. DE--FG02--87ER45334.

**Figure captions:**

`Fig. 1` The low-energy spectrum (a) and PL intensity (b) of excitons confined in a type-II quantum dot as a function of the normal magnetic field for the lowest radial quantum number $n_e = 0$. Insert: Sketch and band diagram for type-II GaSb-GaAs quantum dot.

`Fig. 2` The low-energy spectrum (a) and magnetic moment (b) of excitons confined in a quantum ring as a function of the normal magnetic field in the limit of strong Coulomb interaction. Insert: Sketch of the quantum ring system.